# Tag Clusters as Information Retrieval Interfaces

Kathrin Knautz
Heinrich-Heine-University
Düsseldorf
Dept. of Information Science
Kathrin.Knautz@uni-
duesseldorf.de

Simone Soubusta
Heinrich-Heine-University
Düsseldorf
Dept. of Information Science
Simone.Soubusta@uni-
duesseldorf.de

Wolfgang G. Stock
Heinrich-Heine-University
Düsseldorf
Dept. of Information Science
Stock@phil-fak.uni-
duesseldorf.de

## Abstract

*The paper presents our design of a next generation information retrieval system based on tag co-occurrences and subsequent clustering. We help users getting access to digital data through information visualization in the form of tag clusters. Current problems like the absence of interactivity and semantics between tags or the difficulty of adding additional search arguments are solved. In the evaluation, based upon SERVQUAL and IT systems quality indicators, we found out that tag clusters are perceived as more useful than tag clouds, are much more trustworthy, and are more enjoyable to use.*

## 1. Motivation

The goal of our research project is to develop and evaluate a new information service tool, namely tag clusters (instead of tag clouds), for future web information retrieval systems. Tag clusters make use of digital data (i.e. user-formulated folksonomy-tags) and they can help users find the right access points to documents. Tag clusters represent a new form of information visualization and of visualization-driven query expansion and thus to a new possibility of the application of human-computer interaction (HCI) research in web-based information retrieval (IR).

Today, we are confronted with a lot of so-called Web 2.0 services, e.g. Flickr (photos), YouTube (videos), Technorati (blogs), or Del.icio.us (web pages). Each of these services consists of millions or even billions of documents. The database of Flickr alone encompassed more than 3 billion photos in 2008. Accordingly, the set of suitable matches for more or less unspecific queries will be very large. In order to select small and well-arranged subsets we are in need of methods of query expansion [1, 2], or, more specifically, of query relaxation [3, p. 1840]. The user has to reformulate his query using additional search arguments and (as a general rule) the Boolean operator AND to combine those new query terms with the initial search argument. There are three subtasks: (1) the user must be aware of the option of query expansion; (2) the user has to select correct new terms; and (3) the user has to apply the AND operator correctly.

All three tasks are not unproblematic for internet users. In their famous Excite query log analysis, Spink et al. [4] found huge amounts of problems concerning query reformulation and operator usage. Spink, Jansen and Ozmultu state that their "findings emphasize the need to approach design of Web IR systems, search engines, and even Web site design in a significantly different way than the design of IR systems as practiced to date" [5, p. 327]. As a result of the Excite project they found that only 8% of all queries make use of the Boolean AND (32% incorrectly) and 6% apply the +-sign (39% incorrectly) [6, p. 10, table 4]. A similar study using an AltaVista search log reports that about 80% of all queries do not include any Boolean operator [7, p. 9]. In cases where users apply a Boolean operator they prefer to work with AND. Of Boolean operators "AND was used most. A smaller percentage of queries used OR and a minuscule percentage AND NOT" [4, p. 229]. There are some differences between operator usage in US-based search engines (about 11 – 20%) and European-based search engines (under 5%) [8], but the overall problem seems to be global. The low (correct) usage of Boolean operators in general and of AND in particular can not only be found on web search engines, but also on OPACs (total operator use: 11.7%; AND operator use: 10.3%) [9, p. 1322]. For OPAC the "results show that the failure rate of queries with the 'AND' operator (6.0%) was higher than those of successful searches (4.3%)" [9, p. 1325]. The usage of a professional database, PubMed, is by no means better. Only 11.2% of all queries used at least one (correctly spelled, i.e. with uppercase terms) Boolean operator, and only 10.9% used the AND operator correctly [10, p. 217]. All in all, we can find invalid, incorrect, unsupported or missing operator usage [11, p. 97]. Additionally, users sometimes use incorrect or misspelled search terms [11]. Thus, when people search the web and Web 2.0 services, operator assistance is required [12].

Tag clusters are able to give such assistance:

- following an initial query, they present new search arguments,
- the user can click on every vertex (representing a term) in the cluster (which leads to an automatical ANDing of this term and the initial query),
- the user can click on every edge between two terms in the cluster (which leads to an automatical ANDing of both terms and the initial query).

It is not necessary that the user is aware of the possibility of query relaxation: The system will automatically present the tag cluster of every set of search results. The user can easily select new search arguments: he will find them in the cluster. Finally, the user has no problems with the Boolean AND operator: the system automatically connects the new term (after a click on a vertex) or the new terms (after a click on an edge) with the initial query. The user can repeatedly click on vertices and edges; the system adds all new search arguments to the initial query using AND.

## 2. Tag clouds

A folksonomy [13] is a collection of uncontrolled indexing terms which users unrestrictedly employ to index their individual documents. This form of knowledge organisation stands in stark contrast to nomenclatures, thesauri, and classification systems, which are structural classifications of indexing terms.

As a result of the growing use of tags, a multitude of methods for organising and representing them has sprung up. A lot of websites offer so-called tag clouds as a point of entry for their services. Generally, tag clouds display popular tags in alphabetic order, with the relative size and weight of a tag depending on its relative usage. Thus, tag clouds transform the growing vocabulary of a folksonomy into a navigational tool [14]).

According to Rivadeneira et al. [15], tag clouds perform four functions: With the help of tag clouds, users are able to incorporate tags into their search that represent the sought-after concept, thus facilitating navigation (*search*). They also represent a first point of entry for users with unspecified queries (*browsing*). Moreover, the tag clouds give users an impression of the overall content of the service, a kind of overview (*impression formation/gisting*). Lastly, tag clouds allow for the identification of sets of information that are represented (*recognition*).

Hearst and Rosner [16] observe furthermore that this method of visualization of folksonomies creates a friendly atmosphere and makes the access to a possibly complex service or website easier for first-timers. Moreover, tag clouds allow for the identification of trends through chronological comparison, according to the authors. Russell, amongst others, works on the visualization of the temporal transformation of tag clouds.

Sinclair and Cardew-Hall [18] have examined in how far tag clouds aid in the search for information. They offered users a search box and a tag cloud to conduct their search. They concluded that tag clouds, as visual summaries of content, satisfy all the roles mentioned by Rivadeneira [15]. Additionally, they observed that the process of scanning the cloud and clicking on tags is easier than the formulation of a search query. According to the authors, this facilitates the retrieval in a foreign language [18, p. 27].

Besides these advantages, tag clouds also have several disadvantages. Hearst and Rosner [16, p.2] observe, for example, that the length of a word is merged with its size, thus giving the impression that longer words are more popular.
Furthermore, the depiction of popular tags in the tag cloud implies that those tags have a great semantic density, even though they are often rather useless for purposes of discrimination, owing to their frequent use [19]. Thus, Begelman et al. [20] notice that quite often only a handful of tags and their co-tags dominate a whole tag cloud.

We can infer a further disadvantage from this fact. Since tag clouds usually visualize the whole database, it is quite difficult for new tags – and thus new areas – to be shown at all in a tag cloud. Indentifying trends, as suggested by Hearst and Rosner [16], is thus only possible with a certain time delay and only with the help of a visualization tool like Cloudalicious [17].

Another disadvantage is that tag clouds preclude any form of interaction. The appearance of the tag cloud can only be varied minimally and individual-related aspects cannot be highlighted. Neither can the user customize the tag cloud himself.

However, the greatest drawback of present-day folksonomies, and, consequently, of tag clouds, is the lack of any sort of semantical relations between tags [16] [19]. This drawback is apparent in the discussion of theoretical approaches concerning the structural enhancement of folksonomies, for example, which make frequent use of such terms as "semantic enrichment" [21] and "emergent semantics" [22].

One way to answer the increasing semantics in folksonomies and to support the users of web-based information retrieval systems is to incorporate tag co-occurrences in the visualization. This paper presents the design and implementation of our next generation information retrieval system, which includes the aforementioned aspects and tries to create a user-friendly HCI interface.

## 3. Related Work

The deficits of tag clouds can be countered in two ways: improving the usability of tag clouds [15] [18] [23] [24] [25] or choosing a different kind of visualization to represent the vocabulary. Kaser and Lemire [26], for example, look at different algorithms for optimising the usability of tag clouds, but they are primarily concerned with trying to establish a relation between similar tags. Similarity in this context does not mean that the tags represent the same semantic concept, but rather that they were used to describe the same document. With this approach, Kaser and Lemire took the representation of folksonomy vocabulary in a new direction. They implemented their version of a tag cloud with the help of the classic Knuth-Plass algorithm for the text orientation [27] and based the arrangement on Skiena [28].

As early as 2005, researchers tried to use the similarities between tags, users and links and to display tag clouds in the form of a graph on a map [29].
Begelman, Keller and Smadja [20] are working on an algorithm for the automated clustering of tags on the basis of tag co-occurrences in order to facilitate more effective retrieval. In this approach, the computation of similarities between tags is followed by spectral clustering.

A similar approach is used by Hassan-Montero and Herrero-Solana [19]. Following the computation of tag similarities using the Jaccard similarity coefficient, they clustered the tags hierarchically using the k-means algorithm [30]. The resulting tag cloud displays related tags in relation to each other and is supposed to make the retrieval of related concepts easy. As a result of this restructuring process, unimportant tags have disappeared and frequent tags which are of greater importance and which have a higher degree of discrimination have emerged.

The same approach is used by Cattuto et al. [31] to construct semantic networks on the basis of tag co-occurrences with the goal of comparing the network structures of folksonomies. They are continuing this work in 2009 [32] by analyzing similarities between tags and documents in order to do the growing semantic enrichment justice.

Li et al. [33] demonstrate in how far clustering can facilitate effective browsing in a large set of annotations. Furthermore, they allow for hierarchical and semantical retrieval in the tag cloud generated by their system, ELSABer.

Huang et al. [34] show that a tag cloud generated by cluster algorithms on the basis of tag co-occurrences allows for different viewpoints (e.g. personal, social, universal).

Schrammel et al. [35] evaluated the effects of semantic arrangement versus alphabetical and random arrangement of tags in tag clouds. They observed that a semantically clustered tag cloud with randomly arranged tags yields an improvement for specific queries and aids in directing the users' attention towards tags with a smaller size.

The aforementioned approaches, amongst others [36] [37], are concerned with the growing semantic enrichment in folksonomies and try to present possible solutions.

In addition, several demonstrations of tag visualization are available on the web, five of which are briefly described below.

- Cloudalicious[1] [17] generates graphs in which the assigned tags of a Delicious URL are shown according to their quantitative use in a specific timeframe.
- Extispicious[2] gives you a random textual scattering of a user's tags.
- Netr.it[3] constructs a (manually extendable) co-occurrences network on the basis of one's personal FlickR tags.
- Semantic Cloud[4] generates a semantic tag cloud through clustering on the basis of similarities between tags and moreover offers the possibility of hierarchical retrieval.
- HubLog[5] allows users to visually browse between related Delicious tags.

---

[1] http://cloudalicio.us/
[2] http://kevan.org/extispicious
[3] http://www.netr.it/
[4] http://semanticcloud. rieskamp.info/
[5] http://hublog.hubmed.org/archives/001049.html

## 4. Materials and method

### 4.1 Data Set

Constructing tag clusters requires a set of bookmarks and their tags. We extracted two sets of bookmarks with different topics from the social bookmarking service Delicious, ensuring that both are independent from each other. We designed a parser that extracted the necessary data like URL, tags etc. The first query with the search term "stemmer" resulted in 599 bookmarks. Disregarding the duplicates, we were left with 327 bookmarks and 2,743 tags. A second query with the tags "recipe", "cooking" and "seafood" yielded 684 bookmarks, which were reduced to 518 by removing duplicates, comprising 2,575 tags.

### 4.2 Tag Similarity

The elementary tasks of our approach are the calculation of tag similarities and subsequent clustering. The similarity between two tags can be assessed via different measures and coefficients. The implementation of our tag clusters is based on tag-co-occurrences. Thus we use the common similarity measures Dice (1), Cosine (2) and the Jaccard-Sneath coefficient (3). These measures can be applied to calculate the coincidence-value (φ) for two given tags A and B in the following way:

$$(1) \ \varphi \ (A - B) = \frac{2g}{(a + b)}$$

$$(2) \ \varphi \ (A - B) = \frac{g}{\sqrt{ab}}$$

$$(3) \ \varphi \ (A - B) = \frac{g}{(a + b - g)}$$

Within the framework of folksonomies, the formulas comprise the following variables: a symbolizes all bookmarks containing tag A, b is the number of bookmarks that were tagged with B and g represents bookmarks containing both tags A and B. The possible results range from 0 to 1, where 0 represents no similarity and 1 represents the maximum similarity that can exist between two tags.

In our case, "similarity measures" and "similarity" between tags need to be understood in a figurative sense. We do not state that two terms whose computed values indicate a similarity represent identical concepts, they are not necessarily synonyms. However, we expect them to be meaningfully related in some way. The identification of the exact underlying relation would have to be performed intellectually in a further step. Thus, we can also speak of "coincidence values".

### 4.3 Clustering und Visualization

The next step is the calculation of the coincidence values for all given tag pairs in the database. Afterwards, we can classify the tags by using the computed similarity values. Subsequently, three possible operations can be applied: the single-link, complete-linkage or group-average clustering.

The single-link method starts with the most similar tag pair (A, B). In order to ensure an efficient clustering, this pair of tags has to be contained in 50 bookmarks. Next, we add all tags co-occurring with tag A. A threshold value based on the similarity measure can be used to avoid the cluster's overflow.

Then we add the tags that are similar to tag B, while still considering our threshold value. This step has to be repeated for every new tag until there is not one single tag topping the threshold value left.

```
pseudocode single-link method

for each <tag> in tags:
    if similarity of <start tag 1> and <tag> >= <min value>:
        include <tag> in cluster;
        add edge from <start tag 1> to <tag>

    if similarity of <start tag 2> and <tag> >= <min value>:
        include <tag> in cluster;
        add edge from <start tag 2> to <tag>

for each <tag1> in cluster (increases during runtime):
    for each <tag2> not in cluster:
        if similarity of <tag1> and <tag2> >= <min value>:
            include <tag2> in cluster;
            add edge from <tag1> to <tag2>;
```

By using the complete-linkage method we start again with the most similar tag pair (A, B) while adding only those tags co-occurring with tag A and B in one resource. This step has to be repeated for every new tag gained in this way until there is no tag topping the threshold value left.

```
pseudocode complete-linkage method

for each <tag1> in tags:
    <complete> = true;
    for each <tag2> in tags:
        if <tag2> is in the cluster && similarity of <tag1>
            and <tag2> < <min value>:
            <complete> = false;
            break;

    if <complete>:
        include <tag1> in cluster;

for each <tag1> in cluster:
    for each <tag2> in cluster:
        add edge from <tag1> to <tag2>
```

The group-average method initially operates similar to the single-link method, but after constructing the cluster we calculate an average of the similarity. Then all tags whose similarity with their original tags is below this threshold value are removed from the cluster.

```
pseudocode group-average method

for each <tag> in tags:
    if similarity of <start tag 1> and <tag> >= <min value>:
        include <tag> in cluster;
        <total> = <total> + similarity of <start tag 1> and <tag>;
        <count> = <count> + 1;

    if similarity of <start tag 2> and <tag> >= <min value>:
        include <tag> in cluster;
        <total> = <total> + similarity of <start tag 2> and <tag>;
        <count> = <count> + 1;

for each <tag1> not in cluster (increases during runtime):
    for each <tag2> in cluster:
        if similarity of <tag1> and <tag2> >= <min value>:
            include <tag2> in cluster;
            <total> = <total> + similarity of <tag1> and <tag2>;
            <count> = <count> + 1;

<min value> = <total> / <count>;

if similarity of <start tag 1> and <start tag 2> < <min value>:
    <min value> = similarity of <start tag 1> and <start tag 2>

singleLink();
```

In order to simplify the visualization of the graphs, we adopted the "Java Universal Network/Graph Framework" (JUNG), which is publicized under the BSD licence, for our approach. As we have shown, our tag clusters are based on term-pair-coincidence and consist of undirected graphs. The calculated coincidence-values are visualized by the thickness of the edges between the tags, with ten available line strengths. The font-size represents the quantitative use of the tags, which is already known from common tag clouds. In this case we use a minimum-maximum-normalization to visualize the tags in ten different font-sizes.

### 4.3 Ranking

The query results are presented in two ways. In the first, the bookmarks are ranked by the absolute use of tags. By clicking on an edge between two tags or on another tag the user for all intents and purposes performs a boolean AND operation. In this case the absolute frequency of all tags is accumulated, with the result creating the ranking.

The second alternative makes use of the formula WDF * ITF for a relative ranking. The Within Document Frequency (WDF), which takes the logarithms of the relative occurrences is multiplied with the Inverse Tag- Frequency (ITF) [38], a text statistical value which refers to the total number of tags in the data set.

$$\text{WDF * ITF} = \left[ \frac{ld\ freq(t,b) + 1}{ld\ L} \right] * \left[ \left( ld\ \frac{M}{m} \right) + 1 \right]$$

Freq(t,b) is the occurrence of a tag (t) in a resource (bookmark) (b). That is to say, it represents the frequency of (t). L is the total tag number in a resource (bookmark), namely all tags and their frequency of use. The total number of all tags in the whole folksonomy is M. m is the occurrence of a tag (t) in the set. In our case M and m are related to the initial hit set, because we were not able to calculate the ITF-values from the whole Delicious folksonomy.

### 4.4 User Interface

Despite the complexity of our approach we tried to configure a simple interactive Human-Computer-Interface. We linked all implemented functions with the interface, so that the users have a free choice of all values building up tag clusters. The users are able to use a scroll bar to adjust the threshold (0.- 1.0) and different buttons to set ranking and clustering methods as well as the similarity algorithm.

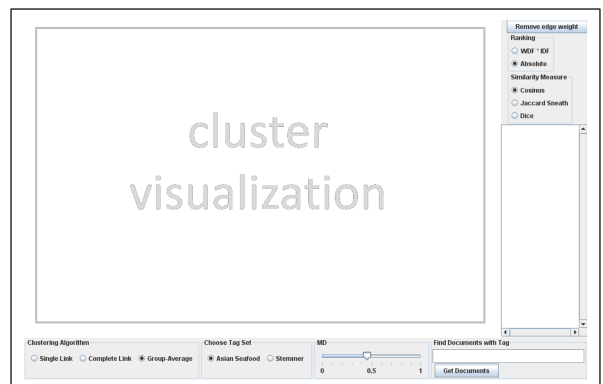

**Figure 1. Applet design of our HCI interface**

### 4.5 Results of clustering

The conception and design of our new information retrieval system do not only give an overview of the content of a database, but also display the results of an initial query using a search box or tag cloud. Through this information visualization the user gains additional support for his research.

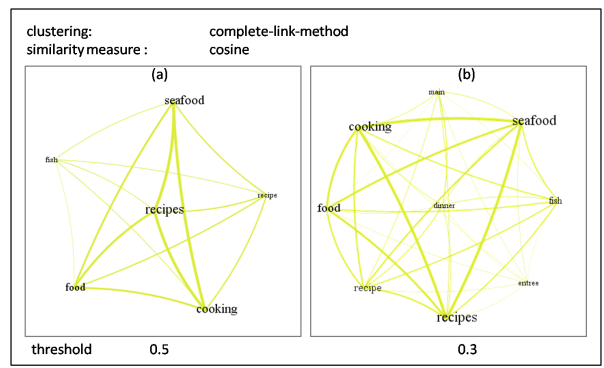

**Figure 2. Automatically generated complete-link-clusters with similarity measure cosine; threshold values 0.5 and 0.3**

Figure 2 displays two clusters built according to the described approach using the complete-linkage method and the cosine similarity measure. The hit number of the initial search was 518 bookmarks containing 2575 tags. The first cluster (a) was built with a threshold of 0.5. The shown syntagmatic net is based on term-pair-coincidences and consists of undirected graphs displaying the calculated coincidence-values. Therefore the similarity of concepts is visualized by the thickness of the edges between the tags. The font-size represents the quantitative use of the tags.

The user is shown a set of results by clicking on a tag in the cluster. If he wants to add other atoms to his search, he can click on the edge between two tags or on other tags. The number of hits changes dynamically. In this way the user can influence his results and is able to select small subsets.

Furthermore, the threshold value of our cluster can be lowered to 0.3. The very specific cluster in (a) could thus be enlarged to yield the one depicted in (b). Every cluster offers access to additional semantically similar tags which users might not have known or thought of at the beginning of their search.

Tag clusters provide the user with access to a graphical/visual retrieval interface. He gets an elaborate alternative to the conventional tag clouds. Syntagmatic networks based on tag co-occurrences are able to deliver the basis for growing demands for semantics within folksonomies. They represent an improvement over the inferior tag clouds, which are no more than a visual summary of folksonomy contents.

# 5. Evaluation

## 5.1 Method

When evaluating concrete IT service tools in Web 2.0 environments such as tag clouds or tag clusters, it does not appear very useful to only work with "classical" indicators of retrieval system quality, i.e. recall and precision. Our IT success model consists of three aspects: the quality of the information system (including its retrieval tool), the quality of the knowledge base, and the quality of the services.

A well-known method for the evaluation of services is SERVQUAL [39]. This method works with two sets of statements, statements to measure expectations about a service category in general (E) and statements to measure perceptions (P) about the category of a particular service. Each statement is accompanied by a seven-point scale ranging from "strongly disagree" (1) to "strongly agree" (7). For each item a difference score Q = P-E is defined. Parasuraman, Zeithaml and Berry [39] defined five service quality dimensions (tangibles, reliability, responsiveness, assurance, and empathy). The assessment is conceptualized as a gap between expectation and perception [40]. It is possible to adopt SERVQUAL for measuring effectiveness of information systems [41]. In IT SERVQUAL, problems arose concerning the exclusive use of the difference score [42] and the pre-defined five quality dimensions [43]. It was discussed to apply not only the difference score, but additional the score of the perceived quality, called SERVPERF [44], or to work exclusively with the perceived-performance-scoring method [45]. If it is handled with caution, SERVQUAL seems to be a valuable tool to measure the IT services quality [46] [47].

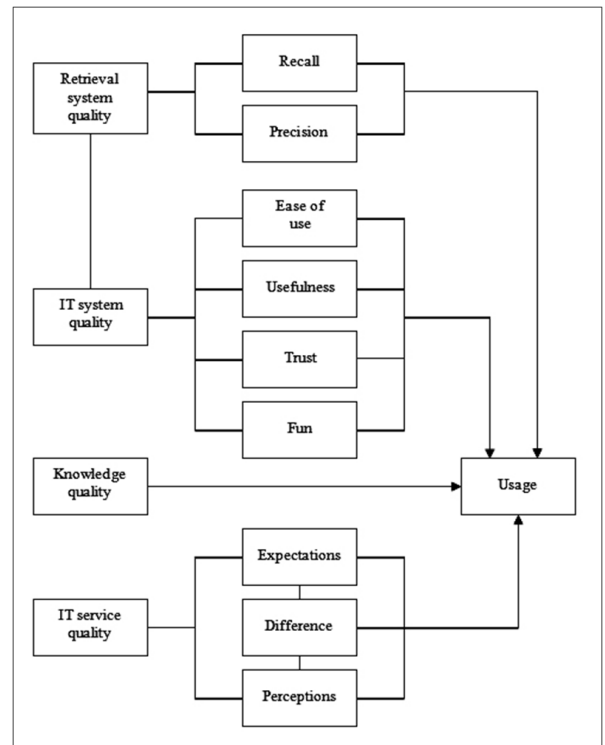

**Figure 3. Our IT success model**

We did not apply pre-defined dimensions, but developed our own dimensions, which are best-fitting to our service tools (tag clouds and clusters). And we calculated all scores, the expectation score, the perception score, and the difference score.

The quality of the IT system component was measured by the classical indicators introduced into IT systems research by Davis, namely perceived usefulness and perceived ease of use [48]. In literature, a further quality indicator [49] is called trust in the information system. Finally, we ask our test persons about their enjoyment when using our tool (often called "perceived playfulness" [50] or "perceived enjoyment" [51]). In our opinion fun is a very important usage aspect in Web 2.0 environments.

In our IT success model you will find additional quality dimensions, namely the knowledge quality (quality of the documents of the information service and quality of the access points to the documents such as folksonomy-tags) and the quality of the information retrieval subsystem (with the established indicators of recall and precision).

In our present study, we are unable to measure those dimensions. The knowledge quality depends on the information service and not on tag clouds or clusters. To measure the quality of a retrieval system you need a lot of queries. Since we worked on our prototype system with only two queries our empirical basis for retrieval evaluation is too small.

Our model of IT success (Fig. 3) is a modified version of the DeLone and McLean model [52] [53] and the Jennex and Olfman model [54].

## 5.2 Results

We made use of a questionnaire to evaluate both the common tag clouds and our new tag cluster tool. We held a pretest with staff members of the Information Science Dept. of the Heinrich-Heine-University Düsseldorf (N = 6) in May 2009. After some light modifications of the questionnaire we asked 30 test persons (students of information science) to evaluate tag clouds (Delicious) and tag clusters in May 2009.

Concerning IT service quality (table 1), the test persons favor tag clusters over tag clouds. There are no great differences between the calculation methods of tag similarity in tag clusters (with a small preference in favor of Cosine). The best performing cluster algorithm seems to be the group average method. We offer two methods of relevance ranking of the result sets and subsets. The WDF*ITF method is perceived as slightly better than ranking by absolute frequency. The test persons have very high expectation scores (and very high perception scores as well) concerning the scroll-bar for adjustment of the similarity threshold value and the line width between the tags in the clusters.

When we look at the IT systems quality evaluation (table 2), we have noticed very significant statistical differences between tag clouds and tag clusters in perceived usefulness, trust and fun, all in favor of tag clusters. Tag clusters are perceived as more useful than tag clouds, are much more trustworthy and significantly more enjoyable. We observed positive correlations between ease of use and all other indicators and of enjoyment and all other indicators (table 3).

**Table 1. Results of evaluation of IT service quality**

| IT service quality indicator | Expectation score (SD) | Perception score (SD) | Difference score |
|---|---|---|---|
| 1. Visualization of folksonomies | | | |
| Tag clouds | 5.69 (0.76) | 3.52 (1.12) | -2.17 |
| Tag clusters | | 5.39 (0.83) | -0.30 |
| 2. Similarity calculations | | | |
| Dice | 5.59 (1.05) | 4.87 (1.36) | -0,72 |
| Jaccard | | 4.73 (1.31) | -0.85 |
| Cosine | | 5.10 (1.06) | -0.49 |
| 3. Cluster algorithms | | | |
| Single link | 5.83 (1.00) | 3.87 (1.50) | -1.96 |
| Complete link | | 4.76 (1.30) | -1.07 |
| Group average | | 5.10 (1.09) | -0.73 |
| 4. Relevance Ranking | | | |
| Absolute frequency | 5.46 (1.29) | 4.39 (1.60) | -1,07 |
| WDF*ITF | | 4.86 (1.43) | -0.61 |
| 5. Tag cluster characteristics | | | |
| Threshold value | 6.07 (0.88) | 5.60 (1.28) | -0.47 |
| Line width | 6.37 (0.85) | 5.55 (1.30) | -0.82 |

N: between 28 and 30; SD: standard deviation; scale: 1 (strongly disagree) to 7 (strongly agree)

Perceived usefulness highly correlates with ease of use (+0.42) and with enjoyment (+0.40). The indicator of enjoyment seems to have a great importance for tag clusters (and maybe for all Web 2.0 tools). In order to explain how this enjoyment is created, we calculated correlation coefficients between enjoyment and all other indicators. We found a high correlation (+0.43) between enjoyment and the expectations for the reasonability of the similarity threshold value adjustment (table 4). The more our test persons expected from the scroll-bar for adjustment of the similarity threshold value, the more these people enjoyed our IT tool (and vice versa).

Obviously, it is enjoyable to change the resolution power of the tag cluster, to add more tags to the cluster (by moving the scroll-bar to the left hand side) and receive a more fine-grained picture of the hit set, or to decrease the number of tags (by moving the scroll-bar more to the right hand side) and see the basic patterns of the tags of the hit set.

**Table 2. Results of evaluation of IT system quality**

| IT systems quality indicator | Tag Clouds Mean (SD) | Tag Clusters Mean (SD) | Significant difference? |
|---|---|---|---|
| Ease of use | 5.17 (1.34) | 5.57 (1.22) | not significant |
| Perceived usefulness | 4.10 (1.35) | 5.66 (1.23) | $\alpha < 0.001$ |
| Trust | 2.90 (1.01) | 4.43 (1.17) | $\alpha < 0.001$ |
| Fun | 4.10 (1.37) | 5.45 (1.35) | $\alpha < 0.001$ |

N: between 28 and 30; SD: standard deviation;
scale: 1 (not at all) to 7 (highly applying)

**Table 3. Correlations (Pearson, two tailed) for the IT systems quality indicators of the tag cluster tool**

|  | Ease of use | Usefulness | Trust | Fun |
|---|---|---|---|---|
| Ease of use | 1 |  |  |  |
| Usefulness | + 0.42 | 1 |  |  |
| Trust | + 0.32 | - 0.08 | 1 |  |
| Fun | + 0.31 | + 0.40 | + 0.19 | 1 |

**Table 4. Correlations (Pearson, two tailed) between the expectation concerning reasonability of the threshold value and the IT systems quality indicators of the tag cluster tool**

| Correlation with the expectation score concerning the reasonability of the threshold value | |
|---|---|
| Ease of use | - 0.01 |
| Usefulness | + 0.16 |
| Trust | + 0.08 |
| Fun | + 0.43 |

## 6. Conclusion and Outlook

The use of syntagmatic relations based on co-occurrences similarity and subsequent clustering provide a lot of opportunities and improvements. Thus the visualization of folksonomies through clustering of tags allows an optimized user access to

necessary digital data on the web. Summing this up, we come to the following points:

- clustering offers a more coherent visual distribution than alphabetical arrangements;
- our approach is a first answer to the growing demands for semantics within folksonomies;
- an equal status of all tags offers the chance that new tags and category groups can be visualized, too. Thus, the missing time-based delimitation can be compensated;
- tag clusters offer the possibility of visualizing even large result sets after an initial search. In this matter the user gains an additional thematic overview of the content;
- the steep structure of tag clouds is dissolved so that users and providers are able to actively interact with the visualization;
- by using tag clusters users are able to adapt the result set to their query. Thus users can independently generate small subsets of all documents relevant for their information need.

Besides an optimization of the loading time of the application, our aim in future work will be the natural language processing (NLP) of tags, like the generalization of number or identification and merging of synonyms through clustering. We are also looking for realizations of sub-clusters through a more differentiated hierarchical clustering.

**Acknowledgements.** We would like to thank the volunteers who participated in the evaluation. Further thanks go to Arkadius Czardybon and Pascal Grün for their support.

## 7. References

[1] E.N. Efthimiadis, "Query expansion", Annual Review of Information Science and Technology, 31, 1996, pp. 121-187.

[2] B.J.Jansen, D.L. Booth, and A. Spink, "Patterns of query reformulation during Web searching", 60(7), 2009, pp. 1358-1371.

[3] G. Kumaran, and J. Allan, "Adapting information retrieval systems to user queries", Information Processing & Management, 44, 2008, pp. 1838-1862.

[4] A. Spink, D. Wolfram, B.J. Jansen, and T. Saracevic, "Searching the Web: The public and their queries", Journal of the American Society for Information Science and Technology, 52(3), 2001, pp. 226-234.

[5] A. Spink, B.J. Jansen, and H.C. Ozmultu, "Use of query reformulation and relevance feedback by Excite users", Internet Research. Electronic Networking Applications and Policy, 10(4), 2000, pp. 317-328.

[6] B.J. Jansen, and A. Spink, J. Bateman, and T. Saracevic, "Real life information retrieval: A study of user queries on the Web", ACM SIGIR Forum, 32(1), 1998, pp. 5-17.

[7] C. Silverstein, M. Henzinger, H. Marais, and M. Moricz, "Analysis of a very large Web search engine query log", ACM SIGIR Forum, 33(1), 1999, pp. 6-12.

[8] B.J. Jansen, and A. Spink, "How are we searching the World Wide Web? A comparison of nine search engine transaction logs", Information Processing & Management, 42, 2006, pp. 248-263.

[9] E.P. Lau, and D.H.L. Goh, "In search of query patterns: A case study of a university OPAC", Information Processing & Management, 42, 2006, pp. 1316-1329.

[10] J.R. Herskovic, L.Y. Tanaka, W. Hersh, and E.V. Bernstam, "A day in the life of PubMed: Analysis of a typical day's query log", Journal of the American Medical Informatics Association, 14(2), 2007, pp. 212-220.

[11] W. Lucas, W., and H. Topi, "Form and function: The impact of query term and operator usage on Web search results", Journal of the American Society for Information Science and Technology, 53(2), 2002, pp. 95-108.

[12] H. Topi, and W. Lucas, "Searching the Web: Operator assistance required", Information Processing & Management, 41, 2005, pp. 383-403.

[13] T. Vander Wal, „Folksonomy", Retrieved May 11, 2009 from http://www.vanderwal.net/folksonomy.html, 2007.

[14] A. Dieberger, P. Dourish, K. Höök, P. Resnick, and A. Wexelblat, "Social Navigation - Techniques for building more usable systems", Interactions, 7, 2000, pp. 36-45.

[15] A.W. Rivadeneira, D.M. Gruen, M.J. Muller, and D.R. Millen, "Getting our head in the clouds: Toward evaluation studies of tagclouds", Proceedings of the SIGCHI Conference on Human Factors in Computing Systems, 2007, pp. 995–998.

[16] M.A. Hearst, and D. Rosner, "Tag clouds: Data analysis tool or social signaller?", Proceedings of 41st Hawaii International Conference on System Sciences (HICSS 2008), Social Spaces minitrack, 2008.

[17] T. Russell, "Cloudalicious: Folksonomy over time", Proceedings of the 6th ACM/IEEEC-CS Joint Conference on Digital Libraries, 2006, p.364.

[18] J. Sinclair, and M. Cardew-Hall, "The folksonomy tag cloud: When is it useful?", Journal of Information Science, 34(1), 2008, pp. 15-29.

[19] Y. Hassan-Montero, and V. Herrero-Solana, "Improving tag clouds as visual information retrieval interfaces", Proceedings of International Conference on Multidisciplinary Information Sciences and Technologies (InSciT2006), Mérida, Spanien, 2006.

[20] G. Begelman, P. Keller, and F. Smadja, "Automated tag clustering: Improving search and exploration in the tag space", Proceedings of the WWW 2006 Collaborative Web Tagging Workshop at WWW2006, Edinburgh, Scotland, 2006.

[21] S. Angeletou, M. Sabou, L. Specia, and E. Motta, "Bridging the gap between folksonomies and the semantic web: An experience report", Workshop: Bridging the Gap between Semantic Web and Web 2.0, European Semantic Web Conference, 2007, pp. 30-43.

[22] L. Zhang, X. Wu, and Y. Yu, "Emergent semantics from folksonomies: A quantitative study", Journal on Data Semantics VI (Lecture Notes in Computer Science), 2006, pp. 168-186.

[23] M. Halvey, and M. Keane, "An assessment of tag presentation techniques", Proceedings of the 16th International Conference on World Wide Web, 2007, pp. 1313–1314.

[24] S. Bateman, C. Gutwin, and M. Nacenta, "Seeing things in the clouds: The effect of visual features on tag cloud selections", Proceedings of the 19th ACM Conference on Hypertext and Hypermedia, Pittsburgh, PA, USA, 2008, pp. 193-202.

[25] S. Seifert, B. Kump, W. Kienreich, G. Granitzer, and M. Granitzer, "On the beauty and usability of tag clouds", Proceedings of the 2008 12th International Conference Information Visualization, Washington, DC, USA: IEEE Computer Society, 2008, pp. 17-25.

[26] O. Kaser, and D. Lemire, "Tag-Cloud drawing: Algorithms", Proc. WWW'07 Workshop on Taggings and Metadata for Social Information Organization, 2007.

[27] D.E. Knuth, and M.F. Plass, "Breaking paragraphs into lines", Software – Practice and Experience, 11 , 1981, pp. 1119–1184.

[28] S. Skiena, The algorithm design manual, Berlin, Springer, 1997.

[29] B. Shaw, "Utilizing folksonomy: Similarity metadata from the Del.icio.us system. Project proposal", Retrieved May 09, 2009 from http://www.metablake.com/webfolk/web-project.pdf, December 2005.

[30] M. Steinbach, G. Karypis, and V. Kumar, "A comparison of document clustering, Proceedings of KDD Workshop on Text Mining, 2000.

[31] C. Cattuto, C. Schmitz, A. Baldassarri, V.D. Servedio, V. Loreto, and A. Hotho, "Network properties of folksonomies", AI Communications, 20(4), 2000, pp. 245-262.

[32] B. Markines, C. Cattuto, F. Menczer, D. Benz, A. Hotho, and G. Stumme "Evaluating similarity measures for emergent semantics of social tagging", Proceedings of the 18th International Conference on World Wide Web, 2009.

[33] R. Li, S. Bao, Y. Yu, B. Fei, and Z. Su, "Towards effective browsing of large scale social annotations", Proceedings of the 16th International Conference on World Wide Web, 2007.

[34] Y.-C. Huang, C.-C Hung, and J.Y. Hsu, "You are what you tag", Proceedings of AAAI 2008 Spring Symposium Series on Social Information Processing, 2008.

[35] J. Schrammel, M. Leitner, and M. Tscheligi, "Semantically structured tag clouds: An empirical evaluation of clustered presentation approaches", Proceedings of the 27th International Conference on Human Factors in Computing Systems, 2009, pp. 2037-2040.

[36] A. Capocci, and G. Caldarelli, "Folksonomies and clustering in the collaborative system CiteULike", Journal of Physics A-Mathematical and Theoretical, 2008, pp. 1-7.

[37] X. Wu, L. Zhang, and Y. Yu, "Exploring social annotations for the semantic web", Proceedings of the 15th International Conference on World Wide Web, 2006, pp. 417-426.

[38] I. Peters, Folksonomies: Indexing and Retrieval on the Web 2.0, (Knowledge and Information. Studies in Information Science, Vol. 1), Saur, Munich, 2009.

[39] A. Parasuraman, V.A. Zeithaml, and L.L. Berry, "SERVQUAL: A multiple-item scale for measuring consumer perceptions of service quality", Journal of Retailing, 64(1), 1988, pp. 12-40.

[40] F. Buttle, "SERVQUAL: Review, critique, research agenda", European Journal of Marketing, 30(1), 1996, pp. 8-32.

[41] L.F. Pitt, R.T. Watson, and C.B. Kavan, "Service quality: A measure of information systems effectiveness", MIS Quarterly, 19(2), 1995, pp. 173-187.

[42] T.P. Van Dyke, L.A. Kappelman, and V.R. Prybutok, "Measuring information systems service quality: Concerns on the use of the SERVQUAL questionnaire", MIS Quarterly, 21(2), 1997, pp. 195-208.

[43] L.F. Pitt, R.T. Watson, and C.B. Kavan, "Measuring information systems quality: Concerns for a complete canvas", MIS Quarterly, 21(2), 1997, pp. 209-221.

[44] W.J. Kettinger, and C.C. Lee, "Pragmatic perspectives on the measurement of information systems service quality", MIS Quarterly, 21(2), 1997, pp. 223-240.

[45] T.P. Van Dyke, V.R. Prybutok, and L.A. Kappelman, "Cautions on the use of the SERVQUAL measure to assess the quality of information systems services", Decision Sciences, 30(3), 1999, pp. 1-15.

[46] W.J. Kettinger, and C.C. Lee, "Replication of measures in information systems research: The case of SERVQUAL", Decision Sciences, 30(3), 1999, pp. 893-899.

[47] J.J. Jiang, G. Klein, and C.L. Carr, "Measuring information system service quality: SERVQUAL from the other side", MIS Quarterly, 26(2), 2002, pp. 145-166.

[48] F.D. Davis, "Perceived usefulness, perceived ease of use, and user acceptance of information technology", MIS Quarterly, 13, 1989, pp. 319-340.

[49] B. Kim, and I. Han, "The role of trust belief and its antecedents in a community-driven knowledge environment", Journal of the American Society for Information Science and Technology, 60(5), 2009, pp. 1012-1026.

[50] C.S. Lin, S. Wu, and R.J. Tsai, "Integrating perceived playfulness into expectation-conformation model for Web portal context", Information and Management, 42(5), 2005, pp. 683-693.

[51] H. van der Heyden, "User acceptance of hedonic information systems", MIS Quarterly, 28(4), 2004, pp. 695-703.

[52] W.H. DeLone, and E.R. McLean, "Information systems success: The quest for the dependent variable", Information Systems Research, 3, 1992, pp. 60-95.

[53] W.H. DeLone, and E.R. McLean, "The DeLone and McLean model of information systems success: A ten-year update", Journal of Management Information Systems, 19(4), 2003, pp. 9-30.

[54] M.E. Jennex, and L. Olfman, L., "A model of knowledge management success", International Journal of Knowledge Management, 2(3), 2006, pp. 51-68.